\pdfoutput=1
\documentclass[english,twocolumn,showpacs,preprintnumbers,amsmath,amssymb,aps,dblfloatfix]{revtex4-1}
\usepackage[T1]{fontenc}
\usepackage[latin9]{inputenc}
\usepackage{color}
\usepackage{array}
\usepackage{multirow}
\usepackage{amssymb}
\usepackage{graphicx}

\makeatletter

\providecommand{\tabularnewline}{\\}

 \@ifundefined{textcolor}{}
 {%
   \definecolor{BLACK}{gray}{0}
   \definecolor{WHITE}{gray}{1}
   \definecolor{RED}{rgb}{1,0,0}
   \definecolor{GREEN}{rgb}{0,1,0}
   \definecolor{BLUE}{rgb}{0,0,1}
   \definecolor{CYAN}{cmyk}{1,0,0,0}
   \definecolor{MAGENTA}{cmyk}{0,1,0,0}
   \definecolor{YELLOW}{cmyk}{0,0,1,0}
 }

\makeatother

\usepackage{babel}
\begin{document}

\title{Beryllium monohydride (BeH): Where we are now, after 86 years of
spectroscopy}

\author{Nikesh S. Dattani,$^{1,2}$}

\email{dattani.nike@gmail.com }

\affiliation{$^{1}$Physical and Theoretical Chemistry Laboratory, Department
of Chemistry, Oxford University, OX1 3QZ, Oxford, UK, }

\affiliation{$^{2}$Quantum Chemistry Laboratory, Department of Chemistry, Kyoto
University, 606-8502, Kyoto, Japan,}
\begin{abstract}
BeH is one of the most important benchmark systems for \emph{ab initio}
methods and for studying Born-Oppenheimer breakdown. However the best
empirical potential and best \emph{ab initio} potential for the ground
electronic state to date give drastically different predictions in
the long-range region beyond which measurements have been made, which
is about $\sim1000$ cm$^{-1}$ for $^{9}$BeH, $\sim3000$ cm$^{-1}$
for $^{9}$BeD, and $\sim13000$ cm$^{-1}$ for $^{9}$BeT. Improved
empirical potentials and Born-Oppenheimer breakdown corrections have
now been built for the ground electronic states $X(1^{2}\Sigma^{+})$
of all three isotopologues. The predicted dissociation energy for
$^{9}$BeH from the new empirical potential is now closer to the current
best \emph{ab initio} prediction by more than 66\% of the discrepancy
between the latter and the previous best empirical potential. The
previous best empirical potential predicted the existence of unobserved
vibrational levels for all three isotopologues, and the current best
\emph{ab initio} study also predicted the existence of all of these
levels, and four more. The present empirical potential agrees with
the \emph{ab initio} prediction of \emph{all} of these extra levels
not predicted by the earlier empirical potential. With one exception,
all energy spacings between vibrational energy levels for which measurements
have been made, are predicted with an agreement of better than 1 cm$^{-1}$
between the new empirical potential and the current best \emph{ab
initio} potential, but some predictions for unobserved levels are
still in great disagreement, and the equilibrium bond lengths are
different by orders of magnitude.
\end{abstract}


\maketitle
With only 5e$^{-}$, BeH is the simplest neutral open shell molecule,
and is therefore of paramount importance in benchmarking \emph{ab
initio} methods. It has been the subject of a plethora of \emph{ab
initio} studies \cite{Cade1967,Chan1968,Bender1969,Jungen1970,Mulliken1971,Popkie1971,Bagus1973,Gerratt1980,Cooper1984,Larsson1984,Larsson1985,Henriet1986,Petsalakis1992,Li1995,Martin1998,Machado1998,Petsalakis1999,Meissner2000,FULSCHER2002,Bruna2003,Bubin2007,Pitarch-Ruiz2008,Pitarch-Ruiz2008a,Koput2011},
with the first Hartree-Fock level endeavor being in 1967. It is also
the second lightest neutral heteronuclear molecule after LiH, and
the only neutral diatomic for which spectroscopic measurements on
a tritium isotopologue have been performed, making it currently the
best benchmark system apart from H$_{2}$ for studying the breakdown
of the Born-Oppenheimer approximation \cite{Martin1998,LeRoy2006a,Bubin2007,Koput2011,Dattani2014a}.
Due to its simplicity, BeH is expected to be present in astronomical
contexts such as exoplanetary atmospheres, cool stars, and the interstellar
medium \cite{Yadin2012}, but in the context of astronomy, has only
been found on our Sun, in the two studies described in \cite{Wohl1971,Shanmugavel2008}.
Finally, the extraordinarily long half-life of the halo nucleonic
atom $^{11}$Be makes $^{11}$BeH a compelling candidate for the formation
of the first halo nucleonic molecule \cite{Dattani2014a}.

Spectroscopic measurements on $^{9}$BeH date back to 1928 \cite{Watson1928,Peterson1928},
and on $^{9}$BeD date back to 1935 \cite{Koontz1935}. By 1937 there
was already an octad of publications on the molecule \cite{Watson1928,Peterson1928,Watson1929,Watson1930,Watson1931,Olsson1932,Koontz1935,Watson1937}.
Since then, higher-resolution spectra have been measured for both
of these isotopologues, and also for $^{9}$BeT in \cite{DeGreef1974}.
Before the present paper, the most thorough empirical analysis of
$^{9}$BeH, $^{9}$BeD, and $^{9}$BeT was that of \cite{LeRoy2006a},
where empirical potentials were built for all three isotopologues,
based on a fit to data from \cite{DeGreef1974,Colin1983,Focsa1998,Shayesteh2003,LeRoy2006a}. 

This state of the art 2006 empirical study left behind various mysteries
which remained unsolved for the last 8 years: 

\begin{figure*}
\caption{\textcolor{black}{\scriptsize \label{fig:BeHwholePotential} Comparison
of the rotationless adiabatic potentials from 2014 {[}this work{]}
and 2011 \cite{Koput2011} for the ground state of }$^{9}$\textcolor{black}{\scriptsize BeH.
Observed vibrational levels are blue and levels predicted by the 2014
potential are gray. The red curve represents the expected long-range
behavior according to theory ($C_{m}$ values are in Table \ref{tab:long-rangeCoefficients}
and damping functions $D_{m}(r)$ are the Douketis-type functions
defined in \cite{LeRoy2011a} with $s=-2$ and $\rho=0.9$).}}

\includegraphics[width=1\textwidth]{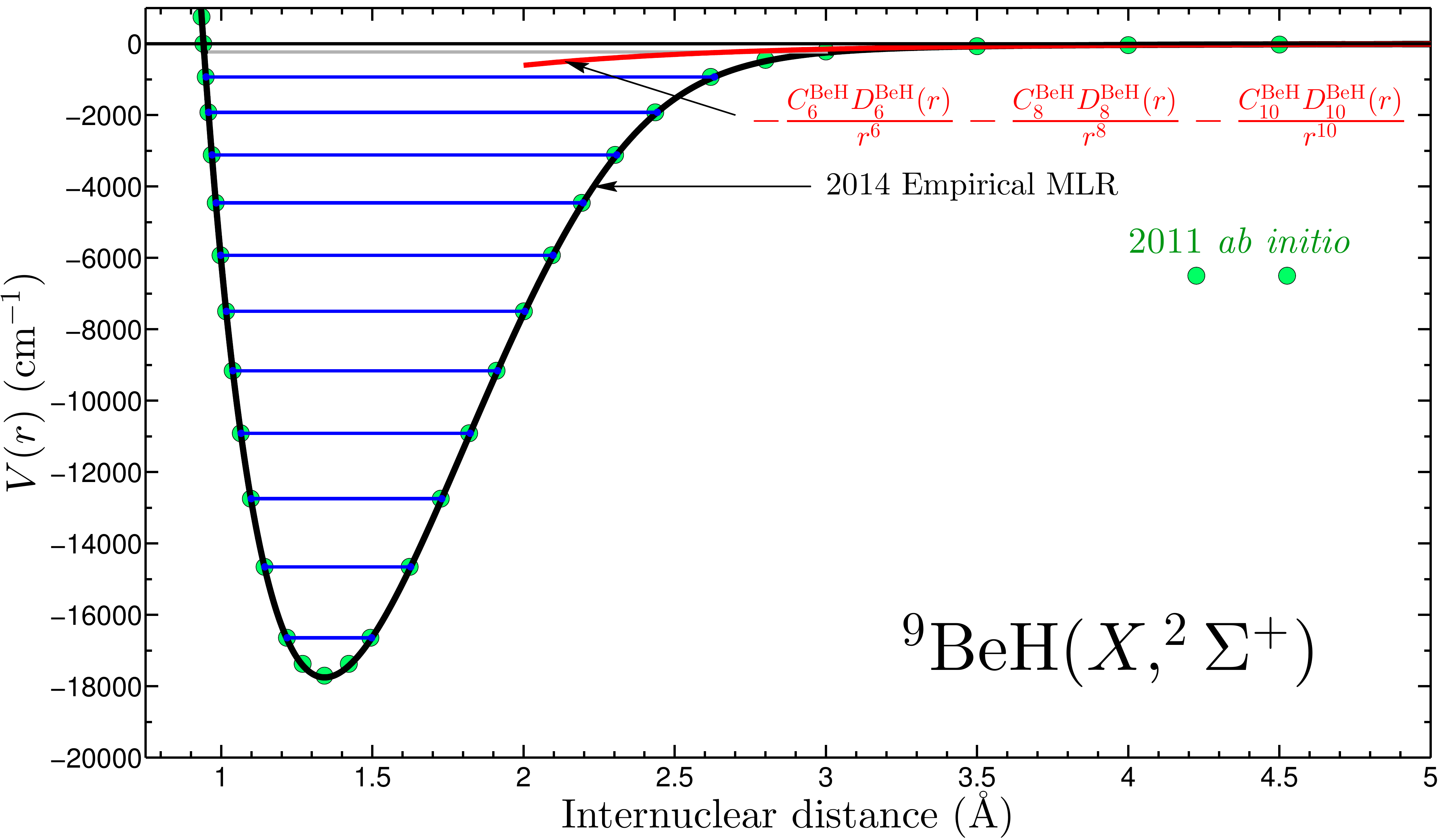}
\end{figure*}

\begin{enumerate}
\item The 2006 study predicted that the $^{9}$BeH dissociation energy was
$\mathfrak{D}_{e}=17590\pm200$ cm$^{-1}$ \cite{LeRoy2006a} which
is higher\emph{ }than the value of $\mathfrak{D}_{e}=17426\pm100$
cm$^{-1}$ in the 1975 experimental study \cite{Colin1975} by more
than the latter's uncertainty. The most recent \emph{ab initio} study
(published in 2011) of $^{9}$BeH \cite{Koput2011} predicted its
dissociation energy to be $\mathfrak{D}_{e}=17702$ cm$^{-1}$ which
is another 112 cm$^{-1}$ higher\emph{ }than the 2006 value, and is
higher than the 1975 value by almost three times the latter's estimated
uncertainty. 
\item The 2006 analysis was unable to determine a meaningful value for the
leading term $u_{0}^{{\rm H}}$ of the adiabatic BOB (Born-Oppenheimer
breakdown) correction function, which helps define the isotopologue
shifts in $\mathfrak{D}_{e}$ and $r_{e}$.
\item The 2006 potentials predicted the existence of 1 more vibrational
level for $^{9}$BeH, 3 more vibrational levels for $^{9}$BeD, and
15 more vibrational levels for $^{9}$BeT than have ever been observed
in experiments. The 2011 \emph{ab initio} study then\emph{ }predicted
the extra levels, but also one further vibrational level for $^{9}$BeH,
2 further levels for $^{9}$BeD, and 1 further vibrational level for
$^{9}$BeT (!). 
\end{enumerate}
As mentioned in the final section of the 2006 paper \cite{LeRoy2006a},
the predictions of the unobserved vibrational levels in that study
were questionable. In fact the large uncertainty in the predicted
value of $\mathfrak{D}_{e}$, the inability to determine a meaningful
value of $u_{0}^{{\rm H}}$, and the lack of confidence in the predictions
of the unobserved vibrational levels were all the result of:
\begin{enumerate}
\item the fact that the highest vibrational level at which data were available,
for any isotopologue, was $\sim1000$ cm$^{-1}$ below the dissociation
limit, and
\item the study used an EMO (expanded Morse oscillator) model for the potential,
which dies off faster than exponentially with respect to the internuclear
distance $r$, while theory dictates that for very large $r$, the
potential should approach the dissociation asymptote with an inverse-power
form (much more slowly than the EMO model does). 
\end{enumerate}
No new data has been published for any of the BeH isotopologues, so
the $\sim1000$ cm$^{-1}$ gap in spectroscopic guidance still presents
a challenge. However, by using a model superior to the EMO we:
\begin{enumerate}
\item are able to bring the $\mathfrak{D}_{e}$ for BeH closer to the 2011
\emph{ab initio} value by more than 66\% of the size of the discrepancy
between the 2006 and 2011 values, 
\item are able to determine $u_{0}^{{\rm H}}$ with an estimated uncertainty
of $\pm10$ cm$^{-1}$,
\item confirm that the extra predicted levels in the 2006 empirical analysis
\cite{LeRoy2006a}, and exactly the same number of further predicted
levels as in the 2011 \emph{ab initio} study \cite{Koput2011} do
in fact exist in our MLR potentials.
\item The present fit to the same spectroscopic data as in 2006, is also
able to reproduce the measured energy transitions on average about
7\% closer than the 2006 potential, with far fewer digits in its parametrization. 
\end{enumerate}
The 2006 study used the EMO model because there was reason to believe
that the long-range tail of the rotationless potential had a barrier
(see discussion and references in the fourth paragraph of this paper's
Conclusions section). However, the 2011 \emph{ab initio} calculation
\cite{Koput2011} has earned our trust (it predicted all but one of
the $^{9}$BeH, $^{9}$BeD, and $^{9}$BeT vibrational spacings in
the data region to within 1 cm$^{-1}$ of the spacings given by the
2006 empirical potential), and it shows no signs of the existence
of a rotationless barrier. Furthermore, the theoretical long-range
potential based on the tremendously accurate long-range constants
($C_{6},$ $C_{8}$, and $C_{10}$) which were recently calculated
\cite{Mitroy2014} for BeH (presented in Table \ref{tab:long-rangeCoefficients}),
also suggests that there is no barrier or turning points in the long-range. 

\begin{table}
\caption{\label{tab:long-rangeCoefficients}Long-range coefficients with and
without finite-mass corrections for hydrogen. $^{\infty}\mbox{H}$
denotes the approximation where a hydrogen atom has an infinite atomic
mass.  All values are in atomic units.  Numbers were generously
provided by Professor Jim Mitroy of Charles Darwin University in (Australia)
\cite{Mitroy2014}. }

\centering{}%
\begin{tabular*}{1\columnwidth}{@{\extracolsep{\fill}}ccccc}
Atom 1 & Atom 2 & $C_{6}$ & $C_{8}$ & $C_{10}$\tabularnewline
\hline 
\noalign{\vskip2mm}
Be & $^{\infty}$H & 34.77918 & 1~213.132 & 46~593.36\tabularnewline
Be & ~~T & 34.79610 & 1~213.825 & 46~624.91\tabularnewline
Be & ~~D & 34.80452 & 1~214.171 & 46~640.62\tabularnewline
Be & ~~H & 34.82984 & 1~215.209 & 46~687.87\tabularnewline
\end{tabular*}
\end{table}

This eliminates any doubt that there may have been, that we should
not use the more accurate MLR (Morse/Long-range) model for the potential,
whose structure is very similar to the EMO for moderate values of
the internuclear distance $r$, but unlike the EMO the correct long-range
form of the potential according to theory, is promised by the mathematical
structure of the MLR. In fact, in 2011 an MLR potential was fitted
to spectroscopic data for the $c(1^{3}\Sigma_{g}^{+})$-state of $^{6,6}$Li$_{2}$
and $^{7,7}$Li$_{2}$, where there was a gap of more than 5000 cm$^{-1}$
between data near the bottom of the potential's well, and data at
the very top \cite{Dattani2011}. In 2013 spectroscopic measurements
were made in the very middle of this gap, and it was found that the
vibrational energies predicted by the MLR potential from \cite{Dattani2011}
were correct to about 1 cm$^{-1}$ \cite{Semczuk2013}. This means
that the MLR model can be capable of making very accurate predictions
outside of the data range, which is particularly pertinent for the
present case of BeH which at the moment still suffers from the issues
mentioned above.

\section{The new potential}

With the exception of using an improved model for the BeH adiabatic
potential energy function $V_{{\rm ad}}^{(1)}(r)$ and the BOB correction
functions $\Delta V_{{\rm ad}}^{(\alpha)}$ and $g^{(\alpha)}(r)$,
we fit the same Hamiltonian to the same data set as used in the 2006
empirical analysis of BeH \cite{LeRoy2006a}. Instead of using the
EMO model of the 2006 study, we use the MLR model for $V_{{\rm ad}}^{(1)}(r)$
, exactly as described for the $a$-state of Li$_{2}$ in 2011\cite{Dattani2011}.
We also use the exact same definitions of $\Delta V_{{\rm ad}}^{(\alpha)}$
and $g^{(\alpha)}(r)$ as in that Li$_{2}$ study \cite{Dattani2011};
the only difference from those used in the 2006 BeH study is the use
of two separate Surkus powers in $\Delta V_{{\rm ad}}^{(\alpha)}$:
$p_{{\rm ad}}$ and $q_{{\rm ad}}$, rather than just $p_{{\rm ad}}$.

For each reference isotopologue, the long-range coefficients were
taken from Table \ref{tab:long-rangeCoefficients}. Relativistic effects
were not taken into account in the calculation of these long-range
coefficients because the relativistic correction for hydrogen is expected
to be about 10 times smaller than the finite-mass correction \cite{Mitroy2014},
and the overall error due to neglecting relativistic effects was expected
to be smaller than the error due to the Be structure model \cite{Mitroy2014}.
Likewise, finite mass corrections for $^{9}$Be were smaller than
the estimated uncertainty in the overall values of the long-range
coefficients \cite{Mitroy2014}. While this was also true for the
isotopes of hydrogen, the \emph{differences} across the three isotopes
\emph{are} expected to be reliable \cite{Mitroy2014}. 

The final MLR potential and BOB correction function parameters are
presented in Table \ref{tab:parametersForPotential}.

\begin{table}
\caption{{\scriptsize Parameters defining the recommended MLR potential for
the $X(^{2}\Sigma^{+})$-state of the reference isotopologue $^{9}$BeH
and the BOB correction functions for all other isotopologues with
$^{9}$Be. Parameters in square brackets were held fixed in the fit,
while numbers in round brackets are 95\% confidence limit uncertainties
in the last digit(s) shown. The potential also incorporates damping
functions according to \cite{LeRoy2011}, with $s=-2$ and $\rho=0.9$.
\label{tab:parametersForPotential}}}

\rule[0.1ex]{1\columnwidth}{0.5pt}

\begin{tabular*}{1\columnwidth}{@{\extracolsep{\fill}}llllll}
\hline 
\noalign{\vskip2mm}
\multicolumn{2}{c}{\textcolor{black}{\footnotesize MLR$_{5,3}^{2.35}(6)$}} & \multicolumn{2}{c}{\textcolor{black}{\footnotesize BOB$_{6,4}^{{\rm ad}}(4)$}} & \multicolumn{2}{c}{\textcolor{black}{\footnotesize BOB$_{3,3}^{{\rm na}}(4,2)$}}\tabularnewline[2mm]
\hline 
\noalign{\vskip2mm}
\textcolor{black}{\footnotesize $\mathfrak{D}_{e}$} & \textcolor{black}{\footnotesize $\;$17664(200) cm$^{-1}$} & \textcolor{black}{\footnotesize $u_{0}^{{\rm H}}$} & \textcolor{black}{\footnotesize $-13(10)$ cm$^{-1}$} & \textcolor{black}{\footnotesize $t_{0}^{{\rm Be}}$} & \textcolor{black}{\footnotesize $[0]$}\tabularnewline
\textcolor{black}{\footnotesize $r_{e}$} & \textcolor{black}{\footnotesize $\;1.342396(2)$ $\mbox{\AA}$} & \textcolor{black}{\footnotesize $u_{1}^{{\rm H}}$} & \textcolor{black}{\footnotesize{} 85.07 cm$^{-1}$} & \textcolor{black}{\footnotesize $t_{1}^{{\rm Be}}$} & \textcolor{black}{\footnotesize $-6.1\times10^{-4}$}\tabularnewline
\textcolor{black}{\footnotesize $C_{6}$} & \textcolor{black}{\footnotesize $\;[34.82984]$ a.u.} & \textcolor{black}{\footnotesize $u_{2}^{{\rm H}}$} & \textcolor{black}{\footnotesize{} 23.51 cm$^{-1}$} & \textcolor{black}{\footnotesize $t_{2}^{{\rm Be}}$} & \textcolor{black}{\footnotesize $5.87\times10^{-3}$}\tabularnewline
\textcolor{black}{\footnotesize $C_{8}$} & \textcolor{black}{\footnotesize $\,[12\,15.209]$ a.u.} & \textcolor{black}{\footnotesize $u_{3}^{{\rm H}}$} & \textcolor{black}{\footnotesize{} 51.10 cm$^{-1}$} & \textcolor{black}{\footnotesize $t_{3}^{{\rm Be}}$} & \textcolor{black}{\footnotesize $-1.2\times10^{-4}$}\tabularnewline
\textcolor{black}{\footnotesize $C_{10}$} & \textcolor{black}{\footnotesize $\,[46\,687.87]$ a.u.} & \textcolor{black}{\footnotesize $u_{4}^{{\rm H}}$} & \textcolor{black}{\footnotesize{} 75.10 cm$^{-1}$} & \textcolor{black}{\footnotesize $t_{4}^{{\rm Be}}$} & \textcolor{black}{\footnotesize $1.05\times10^{-2}$}\tabularnewline
\textcolor{black}{\footnotesize $\beta_{0}$} & \textcolor{black}{\footnotesize{} 1.21862088} & \textcolor{black}{\footnotesize $u_{\infty}^{{\rm H}}$} & \textcolor{black}{\footnotesize $[0]$ cm$^{-1}$} & \textcolor{black}{\footnotesize $t_{\infty}^{{\rm Be}}$} & \textcolor{black}{\footnotesize $[0]$}\tabularnewline
\textcolor{black}{\footnotesize $\beta_{1}$} & \textcolor{black}{\footnotesize{} 1.9636945} &  &  &  & \tabularnewline
\textcolor{black}{\footnotesize $\beta_{2}$} & \textcolor{black}{\footnotesize{} 7.598094} &  &  & \textcolor{black}{\footnotesize $t_{0}^{{\rm H}}$} & \textcolor{black}{\footnotesize $[0]$}\tabularnewline
\textcolor{black}{\footnotesize $\beta_{3}$} & \textcolor{black}{\footnotesize{} 14.16754} &  &  & \textcolor{black}{\footnotesize $t_{1}^{{\rm H}}$} & \textcolor{black}{\footnotesize $1.929\times10^{-3}$}\tabularnewline
\textcolor{black}{\footnotesize $\beta_{4}$} & \textcolor{black}{\footnotesize{} 15.2569} &  &  & \textcolor{black}{\footnotesize $t_{2}^{{\rm H}}$} & \textcolor{black}{\footnotesize $-6.21\times10^{-3}$}\tabularnewline
\textcolor{black}{\footnotesize $\beta_{5}$} & \textcolor{black}{\footnotesize{} 9.2876} &  &  & \textcolor{black}{\footnotesize $t_{\infty}^{{\rm H}}$} & \textcolor{black}{\footnotesize $[0]$}\tabularnewline
\textcolor{black}{\footnotesize $\beta_{6}$} & \textcolor{black}{\footnotesize{} 2.52} &  &  &  & \tabularnewline
\end{tabular*}
\end{table}

\begin{table*}
\caption{\label{tab:comparisonOfSpectroscopicConstants}Comparison of spectroscopic
constants derived from the three studies. Differences denoted by the
columns with a $\Delta$ in the heading are calculated by subtracting
the value in question from the respective 2014 value. The second row
of the header lists the units for each column's quantities.}

\rule[0.1ex]{1\textwidth}{0.5pt}

\begin{tabular*}{1\textwidth}{@{\extracolsep{\fill}}lcccccccc}
\hline 
\noalign{\vskip2mm}
\multirow{2}{*}{} & \textcolor{black}{\small $r_{e}$ } & \textcolor{black}{\small $\Delta r_{e}$ } & \textcolor{black}{\small $\mathfrak{D}_{e}$ } & \textcolor{black}{\small $\Delta\mathfrak{D}_{e}$ } & \textcolor{black}{\small $B_{0}$ } & \textcolor{black}{\small $\Delta B_{0}$ } & \textcolor{black}{\small $D_{0}$$\times10^{-3}$ } & \textcolor{black}{\small $\Delta D_{0}$$\times10^{-3}$ }\tabularnewline
 & \textcolor{black}{\small {[}pm{]}} & \textcolor{black}{\small {[}am{]}} & \textcolor{black}{\small {[}cm$^{-1}${]}} & \textcolor{black}{\small {[}cm$^{-1}${]}} & \textcolor{black}{\small {[}cm$^{-1}${]}} & \textcolor{black}{\small {[}cm$^{-1}${]}} & \textcolor{black}{\small {[}cm$^{-1}${]}} & \textcolor{black}{\small {[}cm$^{-1}${]}}\tabularnewline[2mm]
\hline 
\hline 
\noalign{\vskip2mm}
\multicolumn{9}{c}{\textcolor{black}{\small $^{9}$BeH}}\tabularnewline[2mm]
\hline 
\noalign{\vskip2mm}
\textcolor{black}{\small 2014 Empirical} & \textcolor{black}{\small ~~~~134.239~6(2)} & \textcolor{black}{\small -} & \textcolor{black}{\small ~~~~~~~17~664(200)} & \textcolor{black}{\small -} & \textcolor{black}{\small ~~10.165~63} & \textcolor{black}{\small -} & \textcolor{black}{\small ~1.026~7} & \textcolor{black}{\small -}\tabularnewline
\textcolor{black}{\small 2011 }\textcolor{black}{\emph{\small ab initio}}\textcolor{black}{\small{}
\cite{Koput2011}} & \textcolor{black}{\small 134.144} & 95~600 & \textcolor{black}{\small 17~702} & \textcolor{black}{\small 38} & \textcolor{black}{\small 10.165~0} & \textcolor{black}{\small -0.000~63} & \textcolor{black}{\small 1.027~2} & \textcolor{black}{\small -0.000~5}\tabularnewline
\textcolor{black}{\small 2006 Empirical \cite{LeRoy2006a}} & \textcolor{black}{\small ~~~~~~~134.239~40(12)} & \textcolor{black}{\small 200} & \textcolor{black}{\small ~~~~~~~17~590(200)} & \textcolor{black}{\small 74} & \textcolor{black}{\small ~~10.165~71} & \textcolor{black}{\small -0.000~08} & \textcolor{black}{\small ~1.027~2} & \textcolor{black}{\small -0.000~5}\tabularnewline[2mm]
\hline 
\noalign{\vskip2mm}
\multicolumn{9}{c}{\textcolor{black}{\small $^{9}$BeD}}\tabularnewline[2mm]
\hline 
\noalign{\vskip2mm}
\textcolor{black}{\small 2014 Empirical} & \textcolor{black}{\small ~~~~134.172~8(2)} & \textcolor{black}{\small -} & \textcolor{black}{\small ~~~~~~~17~759(200)} & \textcolor{black}{\small -} & \textcolor{black}{\small 5.625~3} & \textcolor{black}{\small -} & \textcolor{black}{\small 0.313~0} & \textcolor{black}{\small -}\tabularnewline
\textcolor{black}{\small 2011 }\textcolor{black}{\emph{\small ab initio}}\textcolor{black}{\small{}
\cite{Koput2011}} & 134.124 & 48~800 & \textcolor{black}{\small 17~711} & \textcolor{black}{\small 48} & \textcolor{black}{\small 5.626~0} & \textcolor{black}{\small 0.000~7} & \textcolor{black}{\small 0.313~0} & \textcolor{black}{\small 0.000~0}\tabularnewline
\textcolor{black}{\small 2006 Empirical \cite{LeRoy2006a}} & \textcolor{black}{\small ~~~~~~~134.171~97(12)} & \textcolor{black}{\small 8~300} & \textcolor{black}{\small 17~597} & \textcolor{black}{\small 161} & \textcolor{black}{\small 5.625~3} & \textcolor{black}{\small 0.000~0} & \textcolor{black}{\small 0.312~8} & \textcolor{black}{\small 0.000~2}\tabularnewline[2mm]
\hline 
\noalign{\vskip2mm}
\multicolumn{9}{c}{\textcolor{black}{\small $^{9}$BeT}}\tabularnewline[2mm]
\hline 
\noalign{\vskip2mm}
\textcolor{black}{\small 2014 Empirical} & \textcolor{black}{\small ~~~~134.150~4(2)} & \textcolor{black}{\small -} & \textcolor{black}{\small ~~~~~~~177~62(200)} & \textcolor{black}{\small -} & \textcolor{black}{\small 4.106~1} & \textcolor{black}{\small -} & \textcolor{black}{\small 0.166~4} & \textcolor{black}{\small -}\tabularnewline
\textcolor{black}{\small 2011 }\textcolor{black}{\emph{\small ab initio}}\textcolor{black}{\small{}
\cite{Koput2011}} & 134.117 & \textcolor{black}{\small 33~400} & \textcolor{black}{\small 17~715} & \textcolor{black}{\small 47} & \textcolor{black}{\small 4.106~7} & \textcolor{black}{\small -0.000~6} & \textcolor{black}{\small 0.166~4} & \textcolor{black}{\small 0.000~0}\tabularnewline
\textcolor{black}{\small 2006 Empirical \cite{LeRoy2006a}} & \textcolor{black}{\small ~~~~~~~134.149~51(12)} & \textcolor{black}{\small 8~900} & \textcolor{black}{\small 17~599} & \textcolor{black}{\small 163} & \textcolor{black}{\small 4.106~0} & \textcolor{black}{\small -0.000~1} & \textcolor{black}{\small 0.166~3} & \textcolor{black}{\small 0.000~1}\tabularnewline[2mm]
\hline 
\end{tabular*}

\rule[0.1ex]{1\textwidth}{0.5pt}
\end{table*}

\section{Comparison with 2006 empirical and 2011 \emph{ab initio }potentials}

\begin{figure*}
\caption{\textcolor{black}{\scriptsize \label{fig:BeHlongRange} Comparison
of the rotationless adiabatic potentials from 2014 {[}this work{]}
and 2011 \cite{Koput2011} for the ground state of }\textbf{\textcolor{blue}{\scriptsize $^{9}$BeH}}\textcolor{black}{\scriptsize .
Observed vibrational levels are blue and levels predicted by the 2014
potential are gray. The red curve represents the expected long-range
behavior according to theory ($C_{m}$ values are in Table \ref{tab:long-rangeCoefficients}
and damping functions $D_{m}(r)$ are the Douketis-type functions
defined in \cite{LeRoy2011a} with $s=-2$ and $\rho=0.9$).}}

\includegraphics[width=0.95\textwidth]{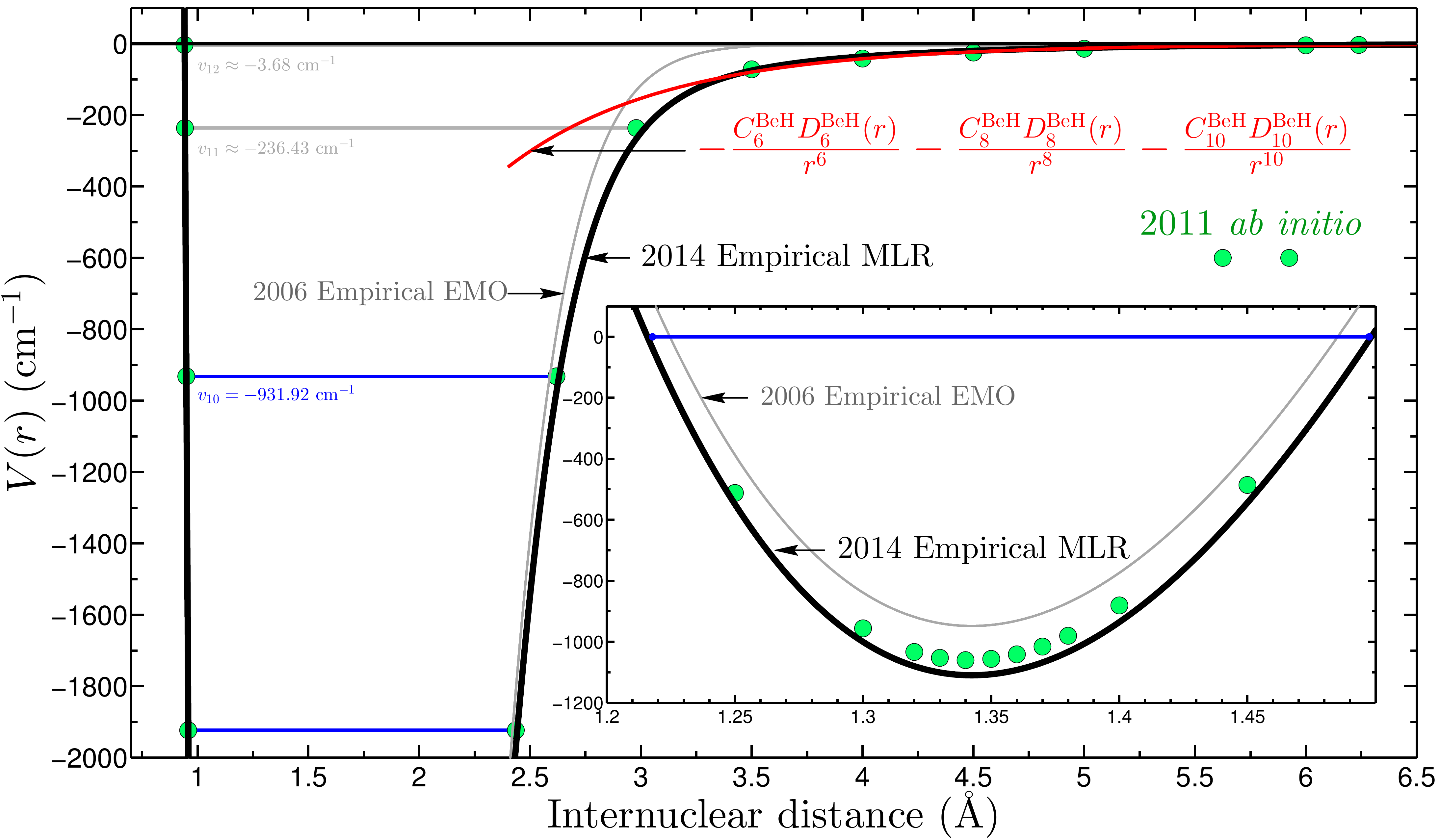}
\end{figure*}

\begin{figure*}
\caption{\textcolor{black}{\scriptsize \label{fig:BeDlongRange} Comparison
of the rotationless adiabatic potentials from 2014 {[}this work{]}
and 2011 \cite{Koput2011} for the ground state of}\textcolor{blue}{\scriptsize{}
}\textbf{\textcolor{blue}{\scriptsize $^{9}$BeD}}\textcolor{black}{\scriptsize .
Observed vibrational levels are blue and levels predicted by the 2014
potential are gray. The red curve represents the expected long-range
behavior according to theory ($C_{m}$ values are in Table \ref{tab:long-rangeCoefficients}
and damping functions $D_{m}(r)$ are the Douketis-type functions
defined in \cite{LeRoy2011a} with $s=-2$ and $\rho=0.9$).}}

\includegraphics[width=0.95\textwidth]{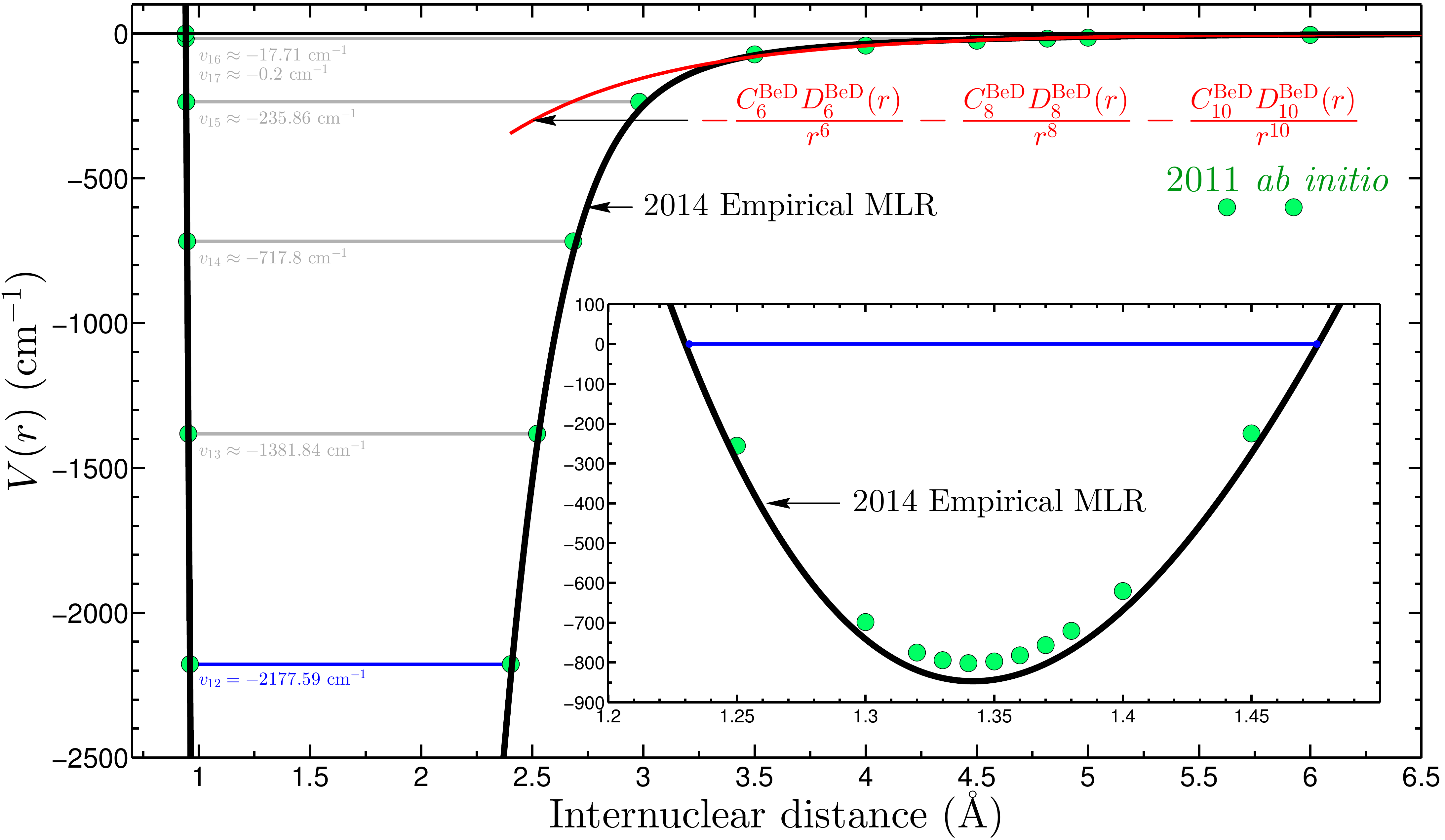}
\end{figure*}

\begin{figure*}
\caption{\textcolor{black}{\scriptsize \label{fig:BeTlongRange} Comparison
of the rotationless adiabatic potentials from 2014 {[}this work{]}
and 2011 \cite{Koput2011} for the ground state of }\textbf{\textcolor{blue}{\scriptsize $^{9}$BeT}}\textcolor{black}{\scriptsize .
Observed vibrational levels are blue and levels predicted by the 2014
potential are gray. The red curve represents the expected long-range
behavior according to theory ($C_{m}$ values are in Table \ref{tab:long-rangeCoefficients}
and damping functions $D_{m}(r)$ are the Douketis-type functions
defined in \cite{LeRoy2011a} with $s=-2$ and $\rho=0.9$).}}

\includegraphics[width=0.95\textwidth]{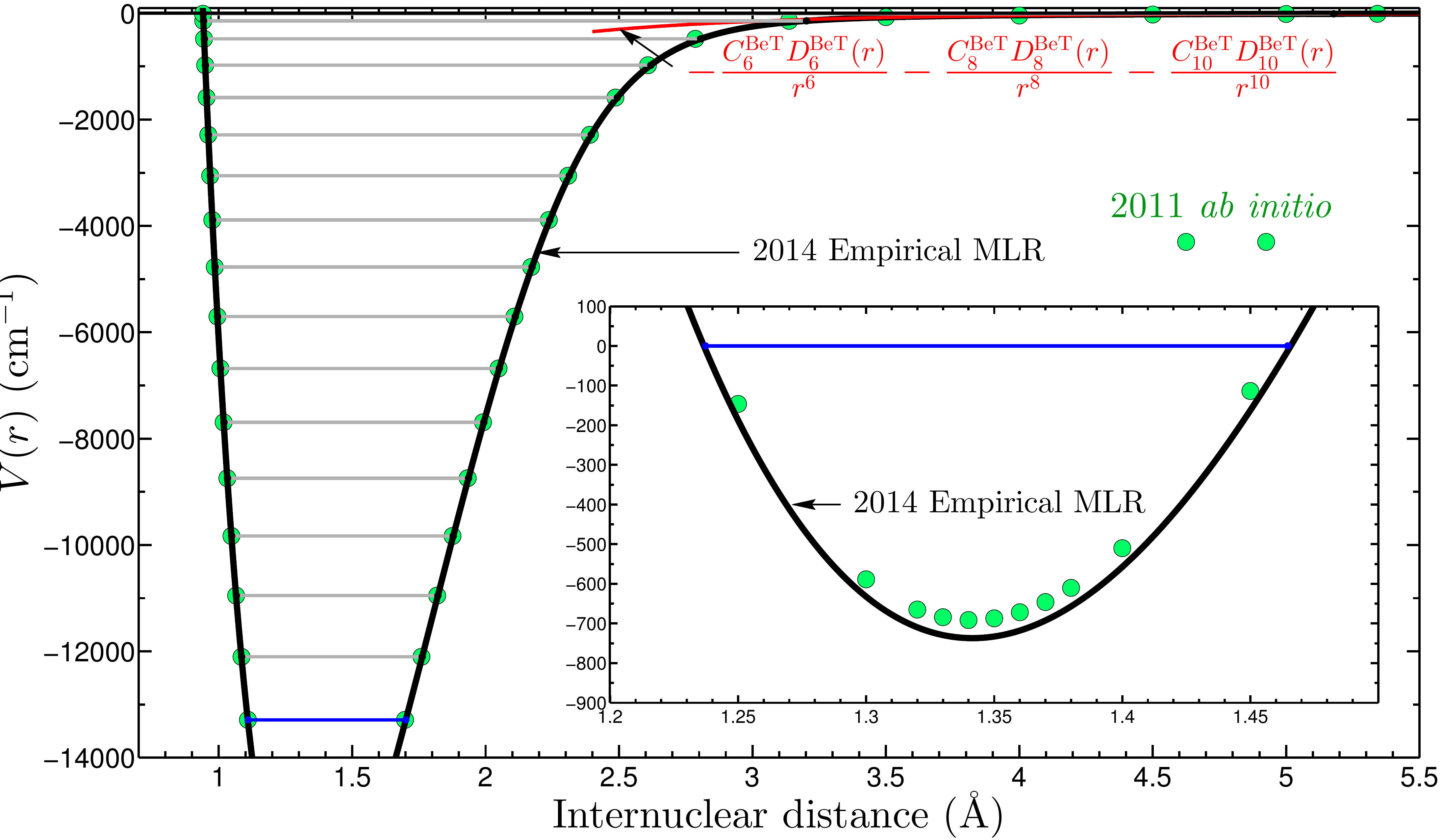}
\end{figure*}

In Figs \ref{fig:BeHwholePotential}, \ref{fig:BeHlongRange}, \ref{fig:BeDlongRange}
and \ref{fig:BeTlongRange} the final adiabatic potentials for all
three isotopologues are compared to the 2006 EMO model results and
to the 2011 \emph{ab initio }results. While \emph{ab initio} potentials
with varying amounts of included theory were discussed in the original
paper \cite{Koput2011}, all figures, tables and discussion about
rotationless potentials in this paper refer to the final adiabatic
potentials for each isotopologue, referred in the original paper as
``CV + F + R + D'' to indicate the use of MR-ACPF/aug-cc-pCV7Z(i)
(denoted by CV), an estimate of electron correlation effects beyond
the approximations of MR-ACPF (denoted by F), second-order Douglas-Kroll-Hess
(DKH) scalar relativistic corrections (denoted by R) and mass-dependent
diagonal BOB corrections (denoted by D). Furthermore, the 2011\emph{
ab inito} study of \cite{Koput2011} also included non-adiabatic BOB
corrections for the calculation of rotational constants $B_{v}$ and
centrifugal distortion constants $D_{v}$. The values of these constants
with the non-adiabatic corrections taken into account were not presented
in the original paper \cite{Koput2011}, but were generously provided
by the author for the present comparison. 

No point-wise representation for these final (CV + F + R + D) potentials
were listed in the original paper \cite{Koput2011}, but they were
generously provided by the author for the present paper. The point-wise\emph{
}representation for each isotopologue listed the (CV + F + R + D)
potential at 48 different internuclear distances, which when all plotted
in Figs \ref{fig:BeHwholePotential}, \ref{fig:BeHlongRange}, \ref{fig:BeDlongRange}
and \ref{fig:BeTlongRange}, proved to make the figures messy. Therefore
these points were only plotted for fairly large values of $r$, and
for small values of $r$, cubic splines were created through the dense
meshes of \emph{ab initio} points, and points were only plotted from
the interpolant at certain places chosen to maximize readability.

\begin{table*}[t]
\caption{{\scriptsize Comparison of the binding energies,  denoted $G(v_{i})$;
zero-point energies (ZPE); and vibrational energy spacings, denoted
$\omega_{i}\equiv G(v_{i})-G(v_{i-1})$; for all three isotopologues.
The last column is the difference between the two columns directly
prior. Discrepancies of $\ge$ 1 cm$^{-1}$ are marked by one star
for spacings outside the data range, and by three stars when the discrepancy
was this large for a spacing  within the data range. Lines with black
font indicate levels that have been observed experimentally and lines
with bold blue font are for unobserved levels that have only been
predicted.\label{tab:ComparisonOfVibrationalEnergies} }}
\rule[0.1ex]{1\textwidth}{0.5pt}

\begin{tabular*}{1\textwidth}{@{\extracolsep{\fill}}cccccccccl}
\hline 
\noalign{\vskip2mm}
\multirow{2}{*}{{\scriptsize $v$}} & {\scriptsize 2006 Empirical} & {\scriptsize 2011 }\emph{\scriptsize ab initio} & {\scriptsize 2014 Empirical } &  & {\scriptsize 2006 Empirical} & {\scriptsize 2011 }\emph{\scriptsize ab initio} & {\scriptsize 2014 Empirical } & \multicolumn{2}{c}{{\scriptsize $\Delta$}}\tabularnewline
 & {\scriptsize{} \cite{LeRoy2006a}} & {\scriptsize{} \cite{Koput2011}} & {\scriptsize {[}Present work{]}} &  & {\scriptsize{} \cite{LeRoy2006a}} & {\scriptsize{} \cite{Koput2011}} & {\scriptsize {[}Present work{]}} & \multicolumn{2}{c}{{\scriptsize {[}$\omega(2014)-\omega(2011)${]}}}\tabularnewline[2mm]
\hline 
\hline 
\noalign{\vskip2mm}
\multicolumn{10}{c}{$^{9}$BeH}\tabularnewline[2mm]
\hline 
\noalign{\vskip2mm}
{\small $0$} & {\small -16567.7651} & {\small -16679.67} & {\small -16641.7725} & {\small ZPE} & {\small 1021.290} & {\small 1022.33} &  &  & \tabularnewline
{\small $1$} & {\small -14581.3475} & {\small -14693.12} & {\small -14655.3569} & {\small $\omega_{1}$} & {\small 1986.4176} & {\small 1986.55} & {\small 1986.4156} & {\small -0.13} & \tabularnewline
{\small $2$} & {\small -12670.8937} & {\small -12782.57} & {\small -12744.9033} & {\small $\omega_{2}$} & {\small 1910.4538} & {\small 1910.55} & {\small 1910.4536} & {\small -0.10} & \tabularnewline
{\small $3$} & {\small -10838.5033} & {\small -10950.10} & {\small -10912.5124} & {\small $\omega_{3}$} & {\small 1832.3904} & {\small 1832.47} & {\small 1832.3909} & {\small -0.08} & \tabularnewline
{\small $4$} & {\small{} -9087.4251} & {\small -9198.99} & {\small -9161.4384} & {\small $\omega_{4}$} & {\small 1751.0782} & {\small 1751.11} & {\small 1751.074} & {\small -0.04} & \tabularnewline
{\small $5$} & {\small{} -7422.5119} & {\small -7534.13} & {\small -7496.5831} & {\small $\omega_{5}$} & {\small 1664.9132} & {\small 1664.86} & {\small 1664.8553} & {\small 0.00} & \tabularnewline
{\small $6$} & {\small{} -5851.1360} & {\small -5962.91} & {\small -5925.3547} & {\small $\omega_{6}$} & {\small 1571.3759} & {\small 1571.22} & {\small 1571.2284} & {\small 0.01} & \tabularnewline
{\small $7$} & {\small{} -4384.7147} & {\small -4496.81} & {\small -4459.1038} & {\small $\omega_{7}$} & {\small 1466.4213} & {\small 1466.10} & {\small 1466.2509} & {\small 0.15} & \tabularnewline
{\small $8$} & {\small{} -3041.1089} & {\small -3153.45} & {\small -3115.5628} & {\small $\omega_{8}$} & {\small 1343.6058} & {\small 1343.36} & {\small 1343.5410} & {\small 0.18} & \tabularnewline
{\small $9$} & {\small{} -1848.9391} & {\small -1962.18} & {\small -1923.3797} & {\small $\omega_{9}$} & {\small 1192.1698} & {\small 1191.27} & {\small 1192.1831} & {\small 0.91} & \tabularnewline
{\small $10$} & {\small{} -857.2981} & {\small -972.95} & {\small -931.9223} & {\small $\omega_{10}$} & {\small{} 991.641} & {\small 989.23} & {\small 991.4574} & {\small 2.23} & {\small {*}{*}{*}}\tabularnewline
\textbf{\textcolor{blue}{\small $11$}} & \textbf{\textcolor{blue}{\small{} -165.6714}} & \textbf{\textcolor{blue}{\small -279.32}} & \textbf{\textcolor{blue}{\small -236.4322}} & \textbf{\textcolor{blue}{\small $\omega_{11}$}} & \textbf{\textcolor{blue}{\small{} 691.6267}} & \textbf{\textcolor{blue}{\small 693.63}} & \textbf{\textcolor{blue}{\small 695.4901}} & \textbf{\textcolor{blue}{\small 1.86}} & \textbf{\textcolor{blue}{\small {*}}}\tabularnewline
\textbf{\textcolor{blue}{\small $12$}} & \textbf{\textcolor{blue}{\small -}} & \textbf{\textcolor{blue}{\small -27.71}} & \textbf{\textcolor{blue}{\small -3.6818}} & \textbf{\textcolor{blue}{\small $\omega_{12}$}} & \textbf{\textcolor{blue}{\small -}} & \textbf{\textcolor{blue}{\small 251.61}} & \textbf{\textcolor{blue}{\small 232.7504}} & \textbf{\textcolor{blue}{\small 18.86}} & \textbf{\textcolor{blue}{\small {*}}}\tabularnewline[2mm]
\hline 
\noalign{\vskip2mm}
\multicolumn{10}{c}{$^{9}$BeD}\tabularnewline[2mm]
\hline 
\noalign{\vskip2mm}
{\small $0$} & {\small -16829.6649} & {\small -16950.59} & {\small -16911.6232} & {\small ZPE} & {\small 759.856} & {\small 760.41} &  &  & \tabularnewline
{\small $1$} & {\small -15340.8226} & {\small -15461.69} & {\small{} -15422.781} & {\small $\omega_{1}$} & {\small{} 1488.8423} & {\small{} 1488.9} & {\small{} 1488.8422} & {\small -0.06} & \tabularnewline
{\small $2$} & {\small -13893.4745} & {\small -14014.32} & {\small -13975.4336} & {\small $\omega_{2}$} & {\small{} 1447.3481} & {\small{} 1447.37} & {\small{} 1447.3474} & {\small -0.02} & \tabularnewline
{\small $3$} & {\small -12488.2889} & {\small -12609.12} & {\small -12570.2475} & {\small $\omega_{3}$} & {\small{} 1405.1856} & {\small{} 1405.2} & {\small{} 1405.1861} & {\small -0.01} & \tabularnewline
{\small $4$} & {\small -11126.2072} & {\small -11247.05} & {\small -11208.1685} & {\small $\omega_{4}$} & {\small{} 1362.0817} & {\small{} 1362.07} & {\small{} 1362.079} & {\small{} 0.01} & \tabularnewline
{\small $5$} & {\small{} -9808.522} & {\small{} -9929.4} & {\small{} -9890.4919} & {\small $\omega_{5}$} & {\small{} 1317.6852} & {\small{} 1317.65} & {\small{} 1317.6766} & {\small{} 0.03} & \tabularnewline
{\small $6$} & {\small{} -8536.9655} & {\small{} -8657.95} & {\small{} -8618.9698} & {\small $\omega_{6}$} & {\small{} 1271.5565} & {\small{} 1271.45} & {\small{} 1271.5221} & {\small{} 0.07} & \tabularnewline
{\small $7$} & {\small{} -7313.8838} & {\small{} -7435.09} & {\small{} -7395.9756} & {\small $\omega_{7}$} & {\small{} 1223.0817} & {\small{} 1222.86} & {\small{} 1222.9942} & {\small{} 0.13} & \tabularnewline
{\small $8$} & {\small{} -6142.5047} & {\small{} -6264.04} & {\small{} -6224.7459} & {\small $\omega_{8}$} & {\small{} 1171.3791} & {\small{} 1171.05} & {\small{} 1171.2297} & {\small{} 0.18} & \tabularnewline
{\small $9$} & {\small{} -5027.3057} & {\small{} -5149.28} & {\small{} -5109.7308} & {\small $\omega_{9}$} & {\small{} 1115.199} & {\small{} 1114.76} & {\small{} 1115.0151} & {\small{} 0.26} & \tabularnewline
{\small $10$} & {\small{} -3974.5183} & {\small{} -4096.98} & {\small{} -4057.1056} & {\small $\omega_{10}$} & {\small{} 1052.7874} & {\small{} 1052.3} & {\small{} 1052.6252} & {\small{} 0.33} & \tabularnewline
{\small $11$} & {\small{} -2992.8739} & {\small{} -3115.86} & {\small{} -3075.5581} & {\small $\omega_{11}$} & {\small{} 981.6444} & {\small{} 981.13} & {\small{} 981.5475} & {\small{} 0.43} & \tabularnewline
\textcolor{black}{\small $12$} & \textcolor{black}{\small{} -2094.8334} & \textcolor{black}{\small{} -2218.65} & \textcolor{black}{\small{} -2177.586} & \textcolor{black}{\small $\omega_{12}$} & \textcolor{black}{\small{} 898.0405} & \textcolor{black}{\small{} 897.21} & \textcolor{black}{\small{} 897.9721} & \textcolor{black}{\small{} 0.76} & \tabularnewline
\textbf{\textcolor{blue}{\small $13$}} & \textbf{\textcolor{blue}{\small{} -1298.8895}} & \textbf{\textcolor{blue}{\small{} -1424.17}} & \textbf{\textcolor{blue}{\small{} -1381.8381}} & \textbf{\textcolor{blue}{\small $\omega_{13}$}} & \textbf{\textcolor{blue}{\small{} 795.9439}} & \textbf{\textcolor{blue}{\small{} 794.48}} & \textbf{\textcolor{blue}{\small{} 795.7479}} & \textbf{\textcolor{blue}{\small{} 1.27}} & \textbf{\textcolor{blue}{\small {*}}}\tabularnewline
\textbf{\textcolor{blue}{\small $14$}} & \textbf{\textcolor{blue}{\small{} -634.5951}} & \textbf{\textcolor{blue}{\small{} -761.77}} & \textbf{\textcolor{blue}{\small{} -717.7958}} & \textbf{\textcolor{blue}{\small $\omega_{14}$}} & \textbf{\textcolor{blue}{\small{} 664.2944}} & \textbf{\textcolor{blue}{\small{} 662.4}} & \textbf{\textcolor{blue}{\small{} 664.0423}} & \textbf{\textcolor{blue}{\small{} 1.64}} & \textbf{\textcolor{blue}{\small {*}}}\tabularnewline
\textbf{\textcolor{blue}{\small $15$}} & \textbf{\textcolor{blue}{\small{} -156.4937}} & \textbf{\textcolor{blue}{\small{} -279.71}} & \textbf{\textcolor{blue}{\small{} -235.8591}} & \textbf{\textcolor{blue}{\small $\omega_{15}$}} & \textbf{\textcolor{blue}{\small{} 478.1014}} & \textbf{\textcolor{blue}{\small{} 482.06}} & \textbf{\textcolor{blue}{\small{} 481.9367}} & \textbf{\textcolor{blue}{\small -0.12}} & \tabularnewline
\textbf{\textcolor{blue}{\small $16$}} & \textbf{\textcolor{blue}{\small -}} & \textbf{\textcolor{blue}{\small{} -50.97}} & \textbf{\textcolor{blue}{\small{} -17.7062}} & \textbf{\textcolor{blue}{\small $\omega_{16}$}} & \textbf{\textcolor{blue}{\small -}} & \textbf{\textcolor{blue}{\small{} 228.75}} & \textbf{\textcolor{blue}{\small{} 218.1529}} & \textbf{\textcolor{blue}{\small -10.6}} & \textbf{\textcolor{blue}{\small {*}}}\tabularnewline
\textbf{\textcolor{blue}{\small $17$}} & \textbf{\textcolor{blue}{\small -}} & \textbf{\textcolor{blue}{\small -8.70}} & \textbf{\textcolor{blue}{\small{} -0.2036}} & \textbf{\textcolor{blue}{\small $\omega_{17}$}} & \textbf{\textcolor{blue}{\small -}} &  & \textbf{\textcolor{blue}{\small{} 17.5026}} & \textbf{\textcolor{blue}{\small 42.27}} & \textbf{\textcolor{blue}{\small {*}}}\tabularnewline[2mm]
\hline 
\noalign{\vskip2mm}
\multicolumn{10}{c}{$^{9}$BeT}\tabularnewline[2mm]
\hline 
\noalign{\vskip2mm}
{\small $0$} & {\small -16940.4368} & {\small -17065.35} & {\small -17025.372} & {\small $\omega_{0}$(ZPE)} & {\small 649.249} & {\small 649.65} &  &  & \tabularnewline
{\small $1$} & {\small -15664.4162} & {\small -15789.30} & {\small -15749.348} & {\small $\omega_{1}$} & {\small 1276.0206} & {\small 1276.05} & {\small 1276.0240} & {\small{} -0.03} & \tabularnewline
{\small $2$} & {\small -14418.5334} & {\small -14543.42} & {\small -14503.465} & {\small $\omega_{2}$} & {\small 1245.8828} & {\small 1245.88} & {\small 1245.8830} & {\small{} 0.00} & \tabularnewline
{\small $3$} & {\small -13203.1649} & {\small -13328.06} & {\small -13288.093} & {\small $\omega_{3}$} & {\small 1215.3685} & {\small 1215.36} & {\small 1215.3720} & {\small{} 0.01} & \tabularnewline
\textbf{\textcolor{blue}{\small $4$}} & \textbf{\textcolor{blue}{\small -12018.8103}} & \textbf{\textcolor{blue}{\small -12143.72}} & \textbf{\textcolor{blue}{\small -12103.738}} & \textbf{\textcolor{blue}{\small $\omega_{4}$}} & \textbf{\textcolor{blue}{\small 1184.3546}} & \textbf{\textcolor{blue}{\small 1184.34}} & \textbf{\textcolor{blue}{\small 1184.3550}} & \textbf{\textcolor{blue}{\small{} 0.02}} & \tabularnewline
\textbf{\textcolor{blue}{\small $5$}} & \textbf{\textcolor{blue}{\small -10866.1397}} & \textbf{\textcolor{blue}{\small -10991.08}} & \textbf{\textcolor{blue}{\small -10951.070}} & \textbf{\textcolor{blue}{\small $\omega_{5}$}} & \textbf{\textcolor{blue}{\small 1152.6706}} & \textbf{\textcolor{blue}{\small 1152.64}} & \textbf{\textcolor{blue}{\small 1152.6680}} & \textbf{\textcolor{blue}{\small{} 0.03}} & \tabularnewline
\textbf{\textcolor{blue}{\small $6$}} & \textbf{\textcolor{blue}{\small{} -9746.0175}} & \textbf{\textcolor{blue}{\small{} -9871.03}} & \textbf{\textcolor{blue}{\small -9830.9584}} & \textbf{\textcolor{blue}{\small $\omega_{6}$}} & \textbf{\textcolor{blue}{\small 1120.1222}} & \textbf{\textcolor{blue}{\small 1120.05}} & \textbf{\textcolor{blue}{\small 1120.1116}} & \textbf{\textcolor{blue}{\small{} 0.06}} & \tabularnewline
\textbf{\textcolor{blue}{\small $7$}} & \textbf{\textcolor{blue}{\small{} -8659.5529}} & \textbf{\textcolor{blue}{\small{} -8784.68}} & \textbf{\textcolor{blue}{\small -8744.5264}} & \textbf{\textcolor{blue}{\small $\omega_{7}$}} & \textbf{\textcolor{blue}{\small 1086.4646}} & \textbf{\textcolor{blue}{\small 1086.35}} & \textbf{\textcolor{blue}{\small 1086.4320}} & \textbf{\textcolor{blue}{\small{} 0.08}} & \tabularnewline
\textbf{\textcolor{blue}{\small $8$}} & \textbf{\textcolor{blue}{\small{} -7608.1843}} & \textbf{\textcolor{blue}{\small{} -7733.52}} & \textbf{\textcolor{blue}{\small -7693.2308}} & \textbf{\textcolor{blue}{\small $\omega_{8}$}} & \textbf{\textcolor{blue}{\small 1051.3686}} & \textbf{\textcolor{blue}{\small 1051.16}} & \textbf{\textcolor{blue}{\small 1051.2956}} & \textbf{\textcolor{blue}{\small{} 0.14}} & \tabularnewline
\textbf{\textcolor{blue}{\small $9$}} & \textbf{\textcolor{blue}{\small{} -6593.8005}} & \textbf{\textcolor{blue}{\small{} -6719.45}} & \textbf{\textcolor{blue}{\small -6678.9691}} & \textbf{\textcolor{blue}{\small $\omega_{9}$}} & \textbf{\textcolor{blue}{\small 1014.3838}} & \textbf{\textcolor{blue}{\small 1014.07}} & \textbf{\textcolor{blue}{\small 1014.2617}} & \textbf{\textcolor{blue}{\small{} 0.19}} & \tabularnewline
\textbf{\textcolor{blue}{\small $10$}} & \textbf{\textcolor{blue}{\small{} -5618.8974}} & \textbf{\textcolor{blue}{\small{} -5744.95}} & \textbf{\textcolor{blue}{\small -5704.2274}} & \textbf{\textcolor{blue}{\small $\omega_{10}$}} & \textbf{\textcolor{blue}{\small{} 974.9031}} & \textbf{\textcolor{blue}{\small{} 974.50}} & \textbf{\textcolor{blue}{\small{} 974.7417}} & \textbf{\textcolor{blue}{\small{} 0.24}} & \tabularnewline
\textbf{\textcolor{blue}{\small $11$}} & \textbf{\textcolor{blue}{\small{} -4686.7826}} & \textbf{\textcolor{blue}{\small{} -4813.32}} & \textbf{\textcolor{blue}{\small -4772.2839}} & \textbf{\textcolor{blue}{\small $\omega_{11}$}} & \textbf{\textcolor{blue}{\small{} 932.1148}} & \textbf{\textcolor{blue}{\small{} 931.63}} & \textbf{\textcolor{blue}{\small{} 931.9435}} & \textbf{\textcolor{blue}{\small{} 0.31}} & \tabularnewline
\textbf{\textcolor{blue}{\small $12$}} & \textbf{\textcolor{blue}{\small{} -3801.8486}} & \textbf{\textcolor{blue}{\small{} -3928.88}} & \textbf{\textcolor{blue}{\small -3887.4944}} & \textbf{\textcolor{blue}{\small $\omega_{12}$}} & \textbf{\textcolor{blue}{\small{} 884.934}} & \textbf{\textcolor{blue}{\small{} 884.44}} & \textbf{\textcolor{blue}{\small{} 884.7895}} & \textbf{\textcolor{blue}{\small{} 0.35}} & \tabularnewline
\textbf{\textcolor{blue}{\small $13$}} & \textbf{\textcolor{blue}{\small{} -2969.9698}} & \textbf{\textcolor{blue}{\small{} -3097.98}} & \textbf{\textcolor{blue}{\small -3055.7135}} & \textbf{\textcolor{blue}{\small $\omega_{13}$}} & \textbf{\textcolor{blue}{\small{} 831.8788}} & \textbf{\textcolor{blue}{\small{} 830.90}} & \textbf{\textcolor{blue}{\small{} 831.7809}} & \textbf{\textcolor{blue}{\small{} 0.88}} & \tabularnewline
\textbf{\textcolor{blue}{\small $14$}} & \textbf{\textcolor{blue}{\small{} -2199.1267}} & \textbf{\textcolor{blue}{\small{} -2327.94}} & \textbf{\textcolor{blue}{\small -2284.9518}} & \textbf{\textcolor{blue}{\small $\omega_{14}$}} & \textbf{\textcolor{blue}{\small{} 770.8431}} & \textbf{\textcolor{blue}{\small{} 770.04}} & \textbf{\textcolor{blue}{\small{} 770.7617}} & \textbf{\textcolor{blue}{\small{} 0.72}} & \tabularnewline
\textbf{\textcolor{blue}{\small $15$}} & \textbf{\textcolor{blue}{\small{} -1500.4838}} & \textbf{\textcolor{blue}{\small{} -1630.46}} & \textbf{\textcolor{blue}{\small -1586.4731}} & \textbf{\textcolor{blue}{\small $\omega_{15}$}} & \textbf{\textcolor{blue}{\small{} 698.6429}} & \textbf{\textcolor{blue}{\small{} 697.48}} & \textbf{\textcolor{blue}{\small{} 698.4787}} & \textbf{\textcolor{blue}{\small{} 1.00}} & \textbf{\textcolor{blue}{\small {*}}}\tabularnewline
\textbf{\textcolor{blue}{\small $16$}} & \textbf{\textcolor{blue}{\small{} -890.4642}} & \textbf{\textcolor{blue}{\small{} -1022.04}} & \textbf{\textcolor{blue}{\small{} -976.768}} & \textbf{\textcolor{blue}{\small $\omega_{16}$}} & \textbf{\textcolor{blue}{\small{} 610.0196}} & \textbf{\textcolor{blue}{\small{} 608.42}} & \textbf{\textcolor{blue}{\small{} 609.7051}} & \textbf{\textcolor{blue}{\small{} 1.29}} & \textbf{\textcolor{blue}{\small {*}}}\tabularnewline
\textbf{\textcolor{blue}{\small $17$}} & \textbf{\textcolor{blue}{\small{} -395.3986}} & \textbf{\textcolor{blue}{\small{} -527.51}} & \textbf{\textcolor{blue}{\small{} -481.3635}} & \textbf{\textcolor{blue}{\small $\omega_{17}$}} & \textbf{\textcolor{blue}{\small{} 495.0656}} & \textbf{\textcolor{blue}{\small{} 494.53}} & \textbf{\textcolor{blue}{\small{} 495.4045}} & \textbf{\textcolor{blue}{\small{} 0.87}} & \tabularnewline
\textbf{\textcolor{blue}{\small $18$}} & \textbf{\textcolor{blue}{\small{} -65.0597}} & \textbf{\textcolor{blue}{\small{} -185.60}} & \textbf{\textcolor{blue}{\small{} -141.9067}} & \textbf{\textcolor{blue}{\small $\omega_{18}$}} & \textbf{\textcolor{blue}{\small{} 330.3389}} & \textbf{\textcolor{blue}{\small{} 341.91}} & \textbf{\textcolor{blue}{\small $339.4568$}} & \textbf{\textcolor{blue}{\small{} -2.45}} & \textbf{\textcolor{blue}{\small {*}}}\tabularnewline
\textbf{\textcolor{blue}{\small $19$}} & \textbf{\small -} & \textbf{\textcolor{blue}{\small{} -40.97 }} & \textbf{\textcolor{blue}{\small{} -9.6258}} & \textbf{\textcolor{blue}{\small $\omega_{19}$}} & \textbf{\textcolor{blue}{\small{} 65.0597}} & \textbf{\textcolor{blue}{\small{} 144.63}} & \textbf{\textcolor{blue}{\small{} 132.2809}} & \textbf{\textcolor{blue}{\small -12.35}} & \textbf{\textcolor{blue}{\small {*}}}\tabularnewline[2mm]
\hline 
\end{tabular*}

\rule[0.1ex]{1\textwidth}{0.5pt}
\end{table*}

Table \ref{tab:comparisonOfSpectroscopicConstants} compares spectroscopic
constants derived from the present study to those derived from the
2006 \cite{LeRoy2006a} and 2011 \cite{Koput2011} studies, and Table
\ref{tab:ComparisonOfVibrationalEnergies} compares the predicted
vibrational binding energies, zero point energies, and vibrational
spacings. A similar comparison was made across Tables IV, V and VI
of \cite{Koput2011}. While those tables did not show $v=12$ for
$^{9}$BeH or $v=17$ for $^{9}$BeD for the 2011 \emph{ab initio}
potentials and $v=13-18$ for $^{9}$BeT for the 2006 empirical potential,
those potentials did in fact support these levels, so their predictions
are included in Table \ref{tab:ComparisonOfVibrationalEnergies} of
the present paper. 

In last column of Table \ref{tab:ComparisonOfVibrationalEnergies},
it is emphasized that the disagreement between the present empirical
potential and the 2011 \emph{ab initio }potential of \cite{Koput2011}
only rises above 1 cm$^{-1}$ for one vibrational spacing in the data
region (the space between $v=9$ and $v=10$ for $^{9}$BeH). This
was also the only vibrational spacing in the data region for which
the disagreement between the predicted energy spacing of the 2011
\emph{ab initio} and the 2006 empirical potential was larger than
1 cm$^{-1}$. However, in the region where measurements have not been
made, the \emph{ab initio} potential is for the most part in much
better agreement with the present empirical potential than the 2006
empirical potential, particularly in the number of vibrational levels
supported and in the energy spacings of higher vibrational energies. 

Furthermore, in Table \ref{tab:comparisonOfSpectroscopicConstants}
we see that the 2011 \emph{ab initio} potentials of \cite{Koput2011}
predicted a dissociation energy $\mathfrak{D}_{e}$ that is 74 cm$^{-1}$
closer to the present empirical potential than the the $\mathfrak{D}_{e}$
predicted by the 2006 empirical potential of \cite{LeRoy2006a}. Combining
this observation with those in the above paragraph, and the comparison
of the potential energy curves in Figs \ref{fig:BeHwholePotential},
\ref{fig:BeHlongRange}, \ref{fig:BeDlongRange} and \ref{fig:BeTlongRange},
tells us that the 2011 \emph{ab initio }potential actually made \emph{more
}accurate predictions than the 2006 empirical potential for various
rotationless properties such as the dissociation energy, vibrational
energy spacings for very high values of $v$, number of predicted
levels, and the potential energy curves themselves are simply in better
agreement.

\emph{}

Finally it is interesting to note that the predicted value for the
leading term of the adiabatic BOB correction function for hydrogen
($u_{0}^{{\rm H}}=-13\pm10$ cm$^{-1}$ in Table \ref{tab:parametersForPotential}),
despite its large uncertainty due to the large gap in data between
the highest observed vibrational level and the dissociation asymptote,
is very similar to the value ($u_{0}^{{\rm H}}=-15.183\pm0.61$ cm$^{-1}$)
obtained for the recent extensive study on the ground ($X,1^{2}\Sigma^{+})$-state
of MgH, for which the fit was to $20\,103$ spectroscopic data across
MgH and MgD (\emph{c.f.} to the present study which involves a fit
to 2821 data across BeH, BeD and BeT).

\section{Conclusion }

Spectroscopic studies of BeH date back to as early as 1928 \cite{Peterson1928,Watson1928,Wigner1928},
and Hartree-Fock level \emph{ab initio }studies on the system go back
as far as 1967 \cite{Cade1967}. At present, about 11\%, 17\%, and
21\% of the adiabatic rotationless potentials for $^{9}$BeH, $^{9}$BeD
and $^{9}$BeT respectively, have not yet been covered by spectroscopic
measurements. Consequently, the dissociation energy $\mathfrak{D}_{e}$,
the number of vibrational levels, and the leading term of hydrogen's
adiabatic BOB correction function $u_{0}^{{\rm H}}$ have been elusive,
the latter having its first empirical value reported in this present
work, and the former two varying greatly in the most recent empirical
\cite{LeRoy2006a} and \emph{ab initio} \cite{Koput2011} studies
before this work.

For all three isotopologues, the confidence intervals for the empirical
values of the $\mathfrak{D}_{e}$'s deduced from the present study
are still about $\pm200$ cm$^{-1}$ which is rather large for a $\mathfrak{D}_{e}$,
especially for such a fundamental system. However, these empirical
values are closer than ever to the state of the art \emph{ab initio}
values \cite{Koput2011} which differ by at most 48 cm$^{-1}$ (in
the case of $^{9}$BeD). 

The present study predicts the \emph{exact} same number of vibrational
levels for all three isotopologues as the state of the art \emph{ab
initio }study \cite{Koput2011}. Of the predicted levels, $v=11-12$
for $^{9}$BeH, $v=13-17$ for $^{9}$BeD, and $v=4-19$ for $^{9}$BeT
have never been observed experimentally.  Of these unobserved transitions,
$v=11$ for $^{9}$BeH, $v=13-15$ for $^{9}$BeD, and $v=4-18$ for
$^{9}$BeT are predicted to be bound by over 100 cm$^{-1}$, by both
the present study and the state of the art \emph{ab initio }study
of \cite{Koput2011}, so they are likely to exist. The higher levels
are in a region where the \emph{ab initio} vibrational energies had
already started to deviate rather greatly from the present empirical
potential, so confirmation of their existence awaits further experiments
or higher level \emph{ab initio }calculations, perhaps with non-adiabatic
BOB corrections included.

Finally, the nature of the long-range tail of BeH potentials has been
an intriguing mystery for several decades. Early electronic structure
studies \cite{Bagus1973,Gerratt1980,Colin1983} discussed the possibility
of there being a rotationless barrier or multiple inflection points
in the long-range region. This was at focus in subsequent \emph{ab
initio} studies which found that these irregularities disappeared
as the basis set size increased or other improvements were made \cite{Cooper1984,Larsson1984,Henriet1986},
and even in later studies which again found such irregularities in
the $r=3-5$\textcolor{black}{{} $\mbox{\AA}$ region \cite{Li1995,Petsalakis1999,Meissner2000,Li1995a}
. Figs \ref{fig:BeHlongRange}, \ref{fig:BeDlongRange} and \ref{fig:BeTlongRange}
provide a compelling reason to believe that no such irregularities
found in earlier }\textcolor{black}{\emph{ab initio }}\textcolor{black}{studies
are}\textcolor{black}{\emph{ }}\textcolor{black}{genuine. The latest
}\textcolor{black}{\emph{ab initio}}\textcolor{black}{{} study of \cite{Koput2011},
which in the data region agrees very well with experiments, did not
show evidence of such irregularities. Since the figures show that
no spectroscopic data is available in the }$r=3-5$\textcolor{black}{{}
$\mbox{\AA}$ region,  the fact that the empirical potentials from
the present analysis also do not appear to have such irregularities
is a consequence of the fact that the model upon which they were based
was designed to smoothly transition between Morse-like short-range
to mid-range behavior, and inverse-power long-range behavior according
to theory. Since the coefficients in Table \ref{tab:long-rangeCoefficients}
are all positive, the theoretical inverse-power tail of the long-range
potential is monotonically decreasing with respect to $r$ and therefore
cannot have a barrier or inflection point. Due to the large gap between
the highest experimentally observed vibrational levels and the point
at which the empirical MLR potentials join the theoretical long-range
potentials (that we see in the figures), there may be some room for
inflections which the MLR model was not able to capture, but the acute
agreement with the accurate}\textcolor{black}{\emph{ ab initio }}\textcolor{black}{potential
energy curves  rules out  the possibility of a barrier and casts
doubt on the existence of inflection points.  }

Finally, the new adiabatic potentials and BOB corrections functions
for $^{9}$BeH, $^{9}$BeD, and $^{9}$BeT presented in this paper
can be used to benchmark \emph{ab initio} methods, especially for
open shell molecules, since BeH is the simplest neutral open shell
molecule with a stable ground electronic state. All of them are provided
for a dense grid of internuclear distance values in a text file in
this paper's Supplementary Material, along with ${\tt MATLAB}$ and
${\tt FORTRAN}$ programs to generate them at any internuclear distance
using their analytic expressions. The present study adds  BeH to
the ever-growing list of molecules for which accurate empirical MLR-type
potentials are available \cite{LeRoy2006,LeRoy2007,Salami2007,Shayesteh2007,Li2008,LeRoy2009,Coxon2010,Stein2010,Li2010,Piticco2010a,LeRoy2011,Ivanova2011,Dattani2011,Xie2011,Yukiya2013,Knockel2013,Semczuk2013,Tritzant-Martinez2013,Wang2013,Li2013a,Li2013,Gunton2013,Meshkov2014,Dattani2014b}.

\section*{Note about references}

While 24 \emph{ab initio} studies were referenced in the introductory
paragraph to this paper, other references may have been missed due
to the author not being aware of their existence. Likewise, experimental
studies before 1937 and after 1974 have been cited, but other studies
about which the author is unaware may exist. If the reader is aware
of any such references, they are keenly encouraged to inform the author
at ${\tt dattani.nike@gmail.com}$.

\section*{Acknowledgments}

\textcolor{black}{It is with pleasure that the author thanks Bob Le~Roy
of University of Waterloo (Canada) for suggesting BeH as an interesting
molecule for testing extrapolation with analytic models. The author
also gratefully thanks Jim Mitroy of Charles Darwin University (Australia)
for providing his unpublished calculated values for the constants
appearing in Table \ref{tab:long-rangeCoefficients} and used for
the models to which the data was fitted, Jacek Koput of Adam Mickiewicz
University (Poland) for his very prompt supply of his pointwise }\textcolor{black}{\emph{ab
initio}}\textcolor{black}{{} potentials for }$^{9}$\textcolor{black}{BeH,
}$^{9}$\textcolor{black}{BeD, and }$^{9}$\textcolor{black}{BeT from
\cite{Koput2011}, Yoshitaka Tanimura of Kyoto University (Japan)
for his generous hospitality, and last but indubitably not least,
Staszek Welsh of Oxford University (UK) for assistance with running
many fits at a preliminary stage of this project. Financial support
was generously provided by JSPS. }

\newpage{}\pagebreak{}\newpage{}

\bibliographystyle{apsrev4-1}

\end{document}